\documentclass{bmvc2k}

\usepackage{graphicx}
\usepackage{xcolor}
\usepackage{amsmath}
\usepackage{bm}
\usepackage{booktabs}

\title{Putting the Segment Anything Model to the Test with 3D Knee MRI - A Comparison with State-of-the-Art Performance}

\addauthor{Oliver Mills}{scojm@leeds.ac.uk}{1}
\addauthor{Nishant Ravikumar}{N.Ravikumar@leeds.ac.uk}{1}
\addauthor{Philip Conaghan}{P.Conaghan@leeds.ac.uk}{1, 2}
\addauthor{Samuel D. Relton}{S.D.Relton@leeds.ac.uk}{1}

\addinstitution{
 University of Leeds\\
 Leeds, UK
}
\addinstitution{
 NIHR Leeds Biomedical Research Centre\\
 Leeds, UK\\
}

\runninghead{Mills et al.}{Putting SAM to the Test with 3D Knee MRI}

\begin{document}

\maketitle

\begin{abstract}
Menisci are cartilaginous tissue found within the knee that contribute to joint lubrication and weight dispersal. Damage to menisci can lead to onset and progression of knee osteoarthritis (OA), a condition that is a leading cause of disability, and for which there are few effective therapies. Accurate automated segmentation of menisci would allow for earlier detection and treatment of meniscal abnormalities, as well as shedding more light on the role the menisci play in OA pathogenesis. Focus in this area has mainly used variants of convolutional networks, but there has been no attempt to utilise recent large vision transformer segmentation models. The Segment Anything Model (SAM) is a so-called foundation segmentation model, which has been found useful across a range of different tasks due to the large volume of data used for training the model. In this study, SAM was adapted to perform fully-automated segmentation of menisci from 3D knee magnetic resonance images. A 3D U-Net was also trained as a baseline. It was found that, when fine-tuning only the decoder, SAM was unable to compete with 3D U-Net, achieving a Dice score of $0.81\pm0.03$, compared to $0.87\pm0.03$, on a held-out test set. When fine-tuning SAM end-to-end, a Dice score of $0.87\pm0.03$ was achieved. The performance of both the end-to-end trained SAM configuration and the 3D U-Net were comparable to the winning Dice score ($0.88\pm0.03$) in the IWOAI Knee MRI Segmentation Challenge 2019. Performance in terms of the Hausdorff Distance showed that both configurations of SAM were inferior to 3D U-Net in matching the meniscus morphology. Results demonstrated that, despite its generalisability, SAM was unable to outperform a basic 3D U-Net in meniscus segmentation, and may not be suitable for similar 3D medical image segmentation tasks also involving fine anatomical structures with low contrast and poorly-defined boundaries.
\end{abstract}

\section{Introduction}
\label{sec:intro}

Osteoarthritis (OA) is one of the leading causes of disability worldwide, and costs the NHS over £3 billion annually \cite{chen_global_2012}. It is a condition where joint cartilage degeneration causes pain and stiffness \cite{buckwalter_impact_2004}. Knee OA is a particularly common type of OA, and with an aging population and rising obesity levels \cite{johnson_epidemiology_2014}, over 8 million people are forecast to have symptomatic knee OA in the UK by 2035 \cite{swain_trends_2020}. Knee OA is a heterogeneous disease \cite{martel-pelletier_magnetic_2023}, and the roles of different tissues in the onset of the disease are not well understood in literature.  The tissues of focus in this study were the menisci, two semi-lunar, wedge-shaped structures found within the knee joint \cite{makris_knee_2011}, which play an important role in load-bearing, as well as the shock distribution and lubrication of the joint \cite{makris_knee_2011, fithian_material_1990}. Multiple studies have shown that meniscal degeneration and tears are highly correlated with the presence of knee osteoarthritis \cite{englund_effect_2007, kornaat_osteoarthritis_2006, bhattacharyya_clinical_2003}, but the role the meniscus plays in the disease pathway is still unclear. A better understanding of the role of the menisci in this regard has the potential to improve early treatment and reduce the burden on health services.

Current methods for assessing meniscal degeneration include `eyeballing' magnetic resonance (MR) scans and arthroscopy. These methods are often ambiguous at assessing the level of meniscal degeneration \cite{rahman_automatic_2020}. Image segmentation of MR scans provides a way to better visualise and analyse the geometric properties of the meniscus \cite{lenchik_automated_2019}. However, manual segmentation is time-consuming and often has poor inter- and intra-individual reliability \cite{mcgrath_manual_2020}. Segmentation of the menisci in MR scans is particularly challenging, as the contrast of the menisci overlaps with other nearby tissues such as femoral and tibial cartilage \cite{rahman_automatic_2020}. Automated segmentation could provide a quicker, objective and more accurate way of segmenting the meniscus \cite{litjens_survey_2017}, and in turn shed more light on the role it plays in OA development. 

\subsection{Deep Learning in Medical Image Analysis}

Since the introduction of convolutional neural networks (CNNs) in 2012 through AlexNet \cite{krizhevsky_imagenet_2012}, which outperformed all other models when classifying the ImageNet data set, CNNs have been the most popular deep learning method for image analysis tasks. CNNs utilise kernels, which are convolution matrices that allow both small and large scale spatial features to be extracted from input images. In the past decade, CNNs have shown great success when applied to a range of medical image analysis tasks \cite{sarvamangala_convolutional_2022, yu_convolutional_2021, ciompi_automatic_2015, menze_multimodal_2015, lotter_multi-scale_2017}. Many applications of CNNs to medical image analysis are now competing with, and even surpassing manual assessment by clinical experts\cite{yu_convolutional_2021}.

In the field of image segmentation, U-Net \cite{ronneberger_u-net_2015} is one of most popular CNN-based methods. The network is made up of two parts: an encoder, where successive convolutional and pooling layers are used to learn a hierarchy of features across spatial scales and reduce the image to a low-dimensional feature representation; and a decoder, where the feature representation is then upscaled to the original image size, often output in the form of a segmentation mask. U-Net also utilises skip-connections, where different stages on the down-sampling path (in the encoder) are concatenated with the corresponding stage on the up-sampling paths (in the decoder) \cite{ronneberger_u-net_2015}. This is done to provide contextual features at the same spatial scale learned in the encoder and reduce information loss during training. U-Net has shown great performance on a range of medical segmentation tasks \cite{siddique_u-net_2021}. A 3D version of U-Net was proposed to allow volumetric segmentation without the need for going slice-wise through a 3D image \cite{cicek_3d_2016}. Currently, variations of the U-Net design are consistently among the top-performing models in both 2D and 3D medical image segmentation challenges \cite{azad_medical_2022, heller_state_2021}.

Recently, image segmentation has been attempted with vision transformer (ViT)-based model architectures, which use attention modules to generate feature embeddings of images. Perhaps the most widely-known example of a ViT segmentation model is the Segment Anything Model (SAM) \cite{kirillov_segment_2023}. SAM was introduced as a foundation 2D image segmentation model, with the ability to generalise to tasks outside the domain of its vast training set, SA-1B, which was made up of over 1.1 million images and 1 billion masks. SAM is also promptable, where point, box, or mask prompts can be provided to help the model generate segmentation masks.   

Since the development of SAM, attention has turned to applying it to medical images. It has been demonstrated that, without any fine-tuning, SAM has the capacity to compete with state-of-the-art segmentation models on certain medical image tasks, albeit only with excessive prompting \cite{deng_segment_2023}. However, other studies have found that SAM struggles in tasks where boundaries are not clearly defined \cite{tang_can_2023}, one example being when attempting skin lesion segmentation \cite{ji_segment_2023}. This is particularly relevant to segmentation of the meniscus, where boundaries are unclear due to the overlap of contrast with neighbouring tissue.

Other studies have tried to fine-tune SAM for medical segmentation tasks \cite{zhang_customized_2023, cheng_sam-med2d_2023, wu_medical_2023}. The first attempt to adapt SAM for medical images fine-tuned the mask decoder on a range of medical images of different modalities, and saw an increase in performance \cite{ma_segment_2023}. However, this study provided prompts during training, so the model was not fully automated. Even after fine-tuning, it was found that SAM struggled with segmentation involving regions that were small, low-contrast, and with unclear boundaries, all of which apply to the meniscus. 

In this study, the aim was to investigate whether SAM could be fine-tuned and adapted to perform automatic segmentation of menisci from 3D knee MR images without providing prompts, which has not yet been attempted. A 3D U-Net was trained using randomly initialised weights on the same data set, to compare SAM to current state-of-the-art performance. The performance of the trained segmentation models was then qualitatively and quantitatively assessed in terms of the similarity between the predicted segmentation masks and the manually generated ground truth masks. 



\vspace{-5pt}
\section{Materials and Methods}
\label{sec:matmeth}

\subsection{Data}

The data used in this study was the same used in the International Workshop on Osteoarthritis Imaging Knee MRI Segmentation Challenge 2019 (IWOAI 2019) \cite{desai_international_2020}.  The data used was a subset of the Osteoarthritis Initiative (OAI).
The OAI contains 3D knee Double-Echo-Steady-State (DESS) MRI of 4796 patients (comprising men and women, aged 45-79) at multiple time-points, who were at risk of femoral-tibial knee osteoarthritis \cite{OAIProtocol, peterfy_osteoarthritis_2008}, making it a valuable resource for studying longitudinal changes of the knee joint. The subset used in this study contained 88 patients at two time points (baseline and 1-year follow-up), each having corresponding manual segmentations of medial and lateral menisci, which were generated by a single expert segmenter from Stryker Imorphics \cite{desai_international_2020}. Each image was 384 by 384, with 160 slices in the sagittal direction. The images had a resolution of (0.365mm $\times$ 0.456mm $\times$ 0.7mm slice thickness) \cite{peterfy_osteoarthritis_2008}.

The 88 patients were split into train, validation and test groups of 60, 14 and 14 respectively, resulting in 176 images split into sets of 120 train, 28 validation, and 28 test images. Splits were done in consistency with the IWOAI 2019 challenge, where Kellgren-Lawrence score, BMI, and sex were approximately equally distributed across the splits \cite{desai_international_2020}. 

\subsection{Preprocessing}

\begin{figure}[t]
\centering
    \includegraphics[width = \linewidth]{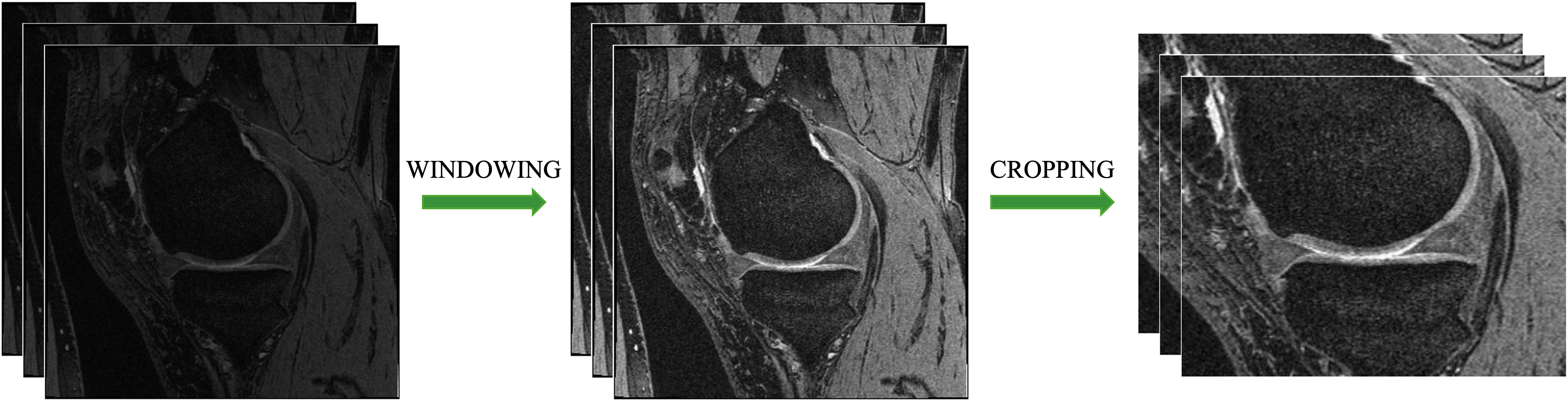}
    \caption{Preprocessing steps performed on the MR Images before model training. Windowing was performed between 0 and 0.005. The cropped region was selected based on the variation in location of ground truths in the train and validation sets.}
    \vspace{-5pt}
    \label{fig:preprocess}
\end{figure}

Before training, windowing was performed on the MRI images, clipping the values to between 0 and 0.005, which were then re-scaled between 0 and 1. This clipping window was selected after viewing the intensity distribution in the images. Clipping the intensity range allowed for greater relative contrast between different artifacts in the joint after re-scaling (Fig.~\ref{fig:preprocess}). The images were then cropped from $384 \times 384 \times 160$ down to $200 \times 256 \times 160$. The cropped region was selected so that all train/validation meniscus masks fell within it, with an extra margin of safety ($\sim20$ voxels) given in all directions. Cropping resulted in the menisci taking up a larger volume of the image, as well as reducing the computational memory requirements of model training. The test images were cropped in the same way.

\subsection{3D U-Net}

The 3D U-Net architecture used took inspiration largely from the original paper that proposed U-Net \cite{ronneberger_u-net_2015}, with 3D convolution operations instead of 2D. The model used 16 feature maps in the first convolution block, with this number doubling in each successive block until the bottleneck. The encoder and decoder were each made up of 3 convolution blocks.

\subsection{Segment Anything Model}

The base ViT version of SAM was used in this project due to the sizeable increase in computational demands from using the ViT Large and ViT Huge versions, which have been reported to provide only marginal improvements in performance in previous studies \cite{ma_segment_2023, kirillov_segment_2023}. To fine-tune SAM, each 3D image was split into 160 separate 2D slices along the sagittal direction, the same being done with the 3D manual segmentations to generate corresponding 2D ground truths. The train, validation and test data splits were kept consistent, so each split now contained the number of 3D images previously mentioned, multiplied by 160. This resulted in train, validation and test splits of size 19200, 4480 and 4480 respectively. Each slice was upsampled using bilinear interpolation and padded. Three copies of this upsampled image were concatenated, resulting in a $1024\times1024\times3$ image in the accepted input format for SAM. When evaluating, all 2D slice predictions were stacked to form a 3D mask, which was compared to the ground truth to evaluate performance. SAM was adapted to not take in prompts and output only a single mask, ensuring that the segmentation was fully automated.

\subsection{Evaluation Metrics}

\subsubsection{Dice Score}

The Dice score is a standard metric for evaluating segmentation performance. Given a ground truth mask, $GT$, and a segmentation prediction, $SP$, the Dice score is given by

\begin{equation}
    Dice = \frac{2 \times |GT \cap SP|}{|GT| + |SP|}.
\end{equation}

\subsubsection{Hausdorff Distance}

The Dice score measures the overlap of two masks, but gives no information on how wrong any false positives are. Two masks could have a high overlap but poor matching on specific structures of interest. To shed light on this, it is useful to include a spatial distance based metric. The Hausdorff Distance is one example of a distance metric. For each point on the surface of the prediction mask $SP$, the distance to the closest point on the surface of $GT$ is measured, and vice versa. The Hausdorff Distance is the maximum of these values, providing information on the worst-matching region of $SP$ to $GT$. However, the maximum Hausdorff Distance can be misleading in the presence of noise and any ground truth outliers \cite{taha_metrics_2015}, so the 95\% percentile Hausdorff Distance was used \cite{huttenlocher_comparing_1993}.

\subsubsection{Average Transverse Thickness}

It has been suggested that the meniscal thickness is a biomarker for Osteoarthritis \cite{dube_where_2018, wirth_three-dimensional_2010}. Therefore, the average transverse thickness of the ground truth and predicted masks were calculated and compared to see how well the predicted masks preserved the thickness of the menisci. This was done by averaging the total volume over the number of non-zero columns in the transverse plane.

\subsection{Model Training}

\begin{table}[b]
    \centering
    \caption{Summary of parameters used for training the different models. SAM was fine-tuned in two ways: by training the mask decoder (SAM 1), and by training end-to-end (SAM 2).}
    \vspace{2mm}
    \begin{tabular}{cccccc}    
    \toprule
        \emph{Model} & \emph{No. of trainable params} & \emph{Batch Size} & \emph{Learning Rate} & \emph{Loss}\\
        \midrule
        SAM 1 & $4,058,340$ & $8$ & $5e-6$ & BCE\\
        SAM 2 & $93,735,472$ & $16$ & $5e-7$ & BCE\\
        3D U-Net & $2,041,825$ & $4$ & $1e-3$ & BCE + Dice\\
        \bottomrule
        \hline
    \end{tabular}
    \label{trainingparams}
\end{table}

SAM was fine-tuned in two different configurations. In the first, the image encoder was frozen and only the mask decoder was trained (SAM 1). This greatly reduced the computational requirements for re-training the model, by lowering the number of trainable parameters in the model from over 91 million down to 4 million. In the second configuration, the model was trained end-to-end with all parameters unfrozen (SAM 2).

The 3D U-Net was trained using a loss function comprising of an unweighted combination of binary cross-entropy (BCE) loss and dice loss, which is simply $1-Dice$. Model convergence was found to be quicker and smoother using this loss function compared to purely BCE or dice loss \cite{jadon_survey_2020}. SAM was trained using BCE loss, due to some slices in the training set containing no ground truth. In this case, dice loss would become large, due to no overlap, and would lead to unstable training. The Adam optimiser was used for training both the 3D U-Net and SAM. Random grid searches were performed to select the optimal batch size and learning rate (as well as number of kernels for 3D U-Net) for each of the models (Table~\ref{trainingparams}). Training was stopped once the validation loss failed to decrease for 5 epochs.

\begin{table}[t]
    \centering
    \caption{A comparison of model performance metrics on the test set. Average Thickness Difference was calculated by taking the mean of the predicted mask thickness subtracted from the ground truth thickness for each test case. SAM was fine-tuned both by training only the mask decoder (SAM 1), and by training end-to-end (SAM 2). Hausdorff Distance and Average Thickness Difference are reported in millimeters (mm).}
    \vspace{2mm}
    \begin{tabular}{cccc}    
    \toprule
        \emph{Model} & \emph{Dice score} & \emph{Hausdorff Distance} & \emph{Average Thickness Difference}\\
        \midrule
        SAM 1 & $0.81\pm0.03$ & $3.1\pm1.9$ & $-0.17\pm0.2$\\
        SAM 2 & $\bm{0.87\pm0.03}$ & $2.4\pm1.4$ & $0.07\pm0.12$ \\
        3D U-Net & $\bm{0.87\pm0.03}$ & $\bm{1.8\pm0.8}$ & $\bm{0.03\pm0.15}$ \\
        \bottomrule
        \hline
    \end{tabular}
    \label{modelcomparison}
    \vspace{-5pt}
\end{table}

\section{Results and Discussion}

SAM 1 achieved the worst score on all metrics (Table~\ref{modelcomparison}), suggesting that, despite SAM's large pre-trained encoder, the extracted features were not good enough to produce competitive meniscus segmentations. Training SAM end-to-end improved Dice score performance, matching 3D U-Net. In the IWOAI 2019 challenge, the highest-performing entry achieved a Dice score of $0.88\pm0.3$ on the test set \cite{desai_international_2020}, meaning that SAM 2 was able to compete with state-of-the-art performance despite lacking 3D context. This also demonstrated the impressive segmentation ability of the vanilla 3D U-Net, achieving a similarly high score.

\begin{figure}[b!]
\centering
    \begin{subfigure}[h]{0.48\linewidth}
    \centering
        \includegraphics[width = \linewidth]{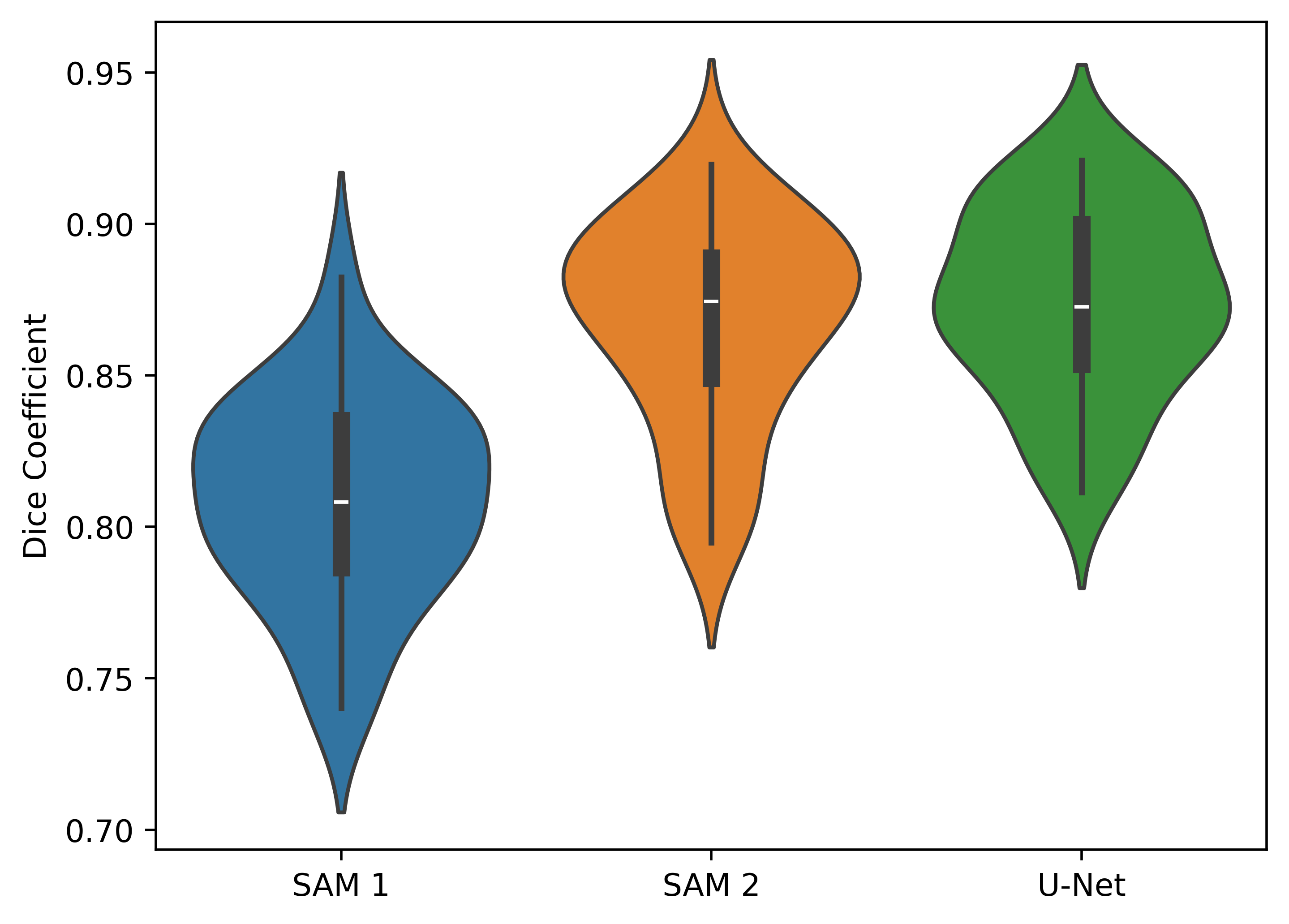}
        \caption{Dice score distributions}
        \label{fig:dice}
    \end{subfigure}
    \hfill
    \begin{subfigure}[h]{0.48\linewidth}
    \centering
        \includegraphics[width = \linewidth]{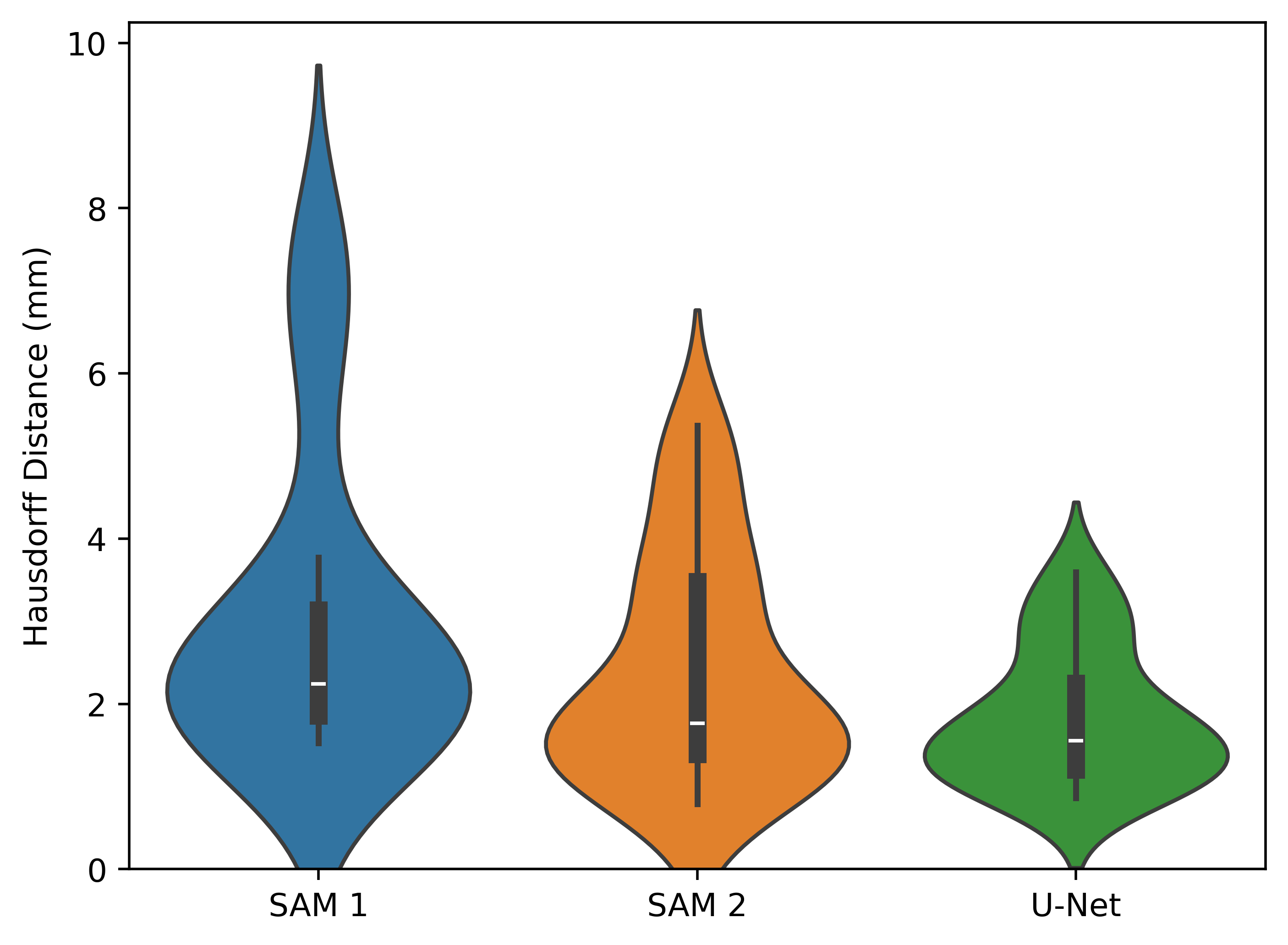}
        \caption{Hausdorff distance distributions}
        \label{fig:hausdorff}
    \end{subfigure}
    \vspace{5pt}
    \caption{Violin plots showing the distributions of the dice score (a) and Hausdorff distance (b) of the three model configurations when predicting on the test set. In the violin interior, a box plot is shown. In SAM 1, only the mask decoder was trained. In SAM 2, the model was trained end-to-end.}
    \label{fig:violins}
\end{figure}

There was little to separate SAM 2 and 3D U-Net on mean values alone, so the distributions of the Dice score and Hausdorff Distance across the test set were plotted (Fig.~\ref{fig:violins}). When looking at Dice Score, SAM 2 and U-Net had a similar mean and interquartile range, but SAM 2 performed more poorly on a small number of cases (\ref{fig:dice}). Fig.~\ref{fig:hausdorff} indicates that despite the improvement from training SAM end-to-end compared to training only the decoder, 3D U-Net has both a lower mean and less variability than both SAM configurations in terms of the Hausdorff distance. This suggests that 3D U-Net is superior at predicting masks that closely match the geometry of the menisci consistently, without introducing errant spatial artifacts.

\begin{figure}[t!]
\centering
    \begin{subfigure}[h]{0.48\linewidth}
    \centering
        \includegraphics[width = \linewidth]{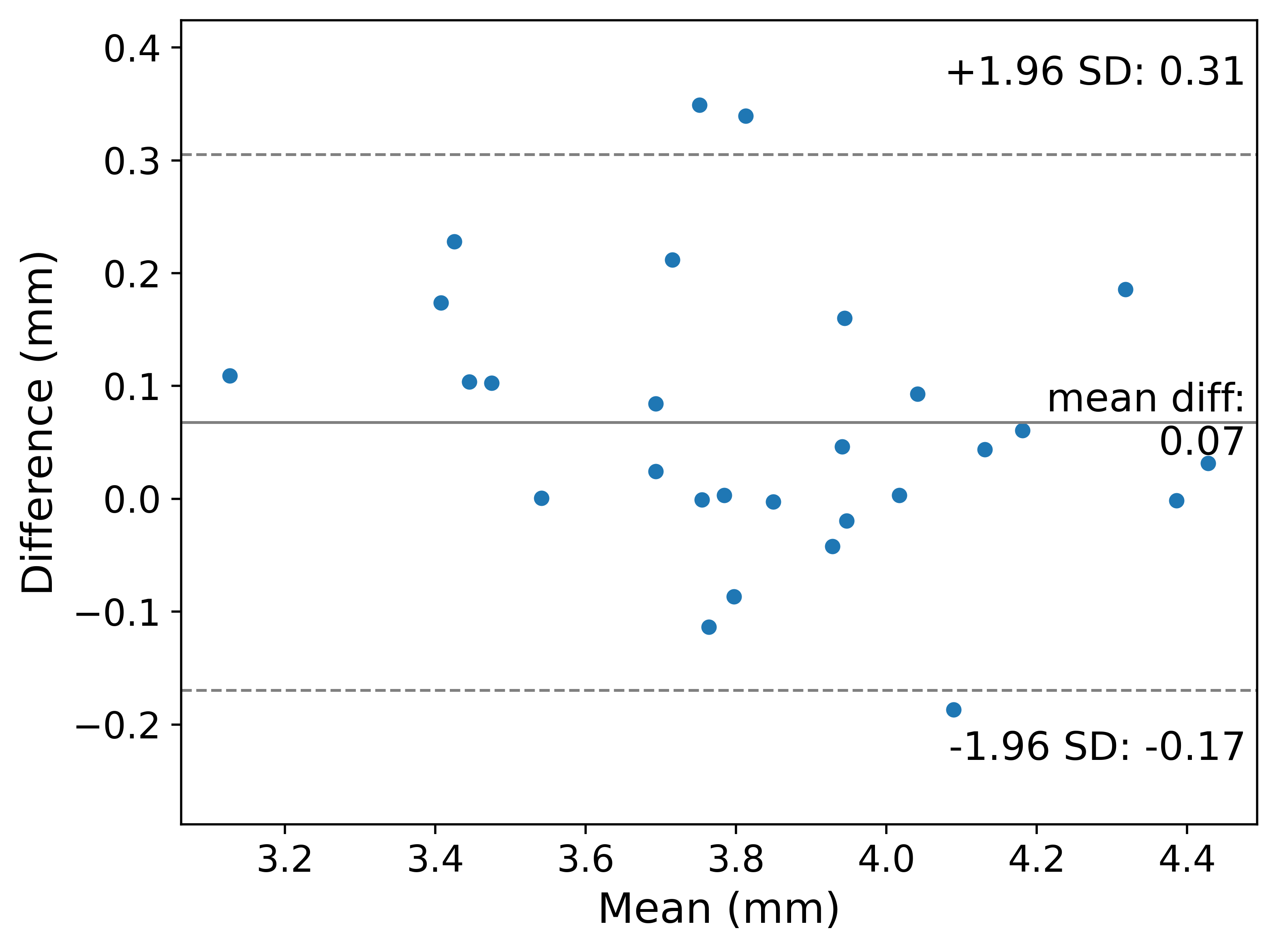}
        \caption{SAM 2}
        \label{fig:BAsam}
    \end{subfigure}
    \hfill
    \begin{subfigure}[h]{0.48\linewidth}
    \centering
        \includegraphics[width = \linewidth]{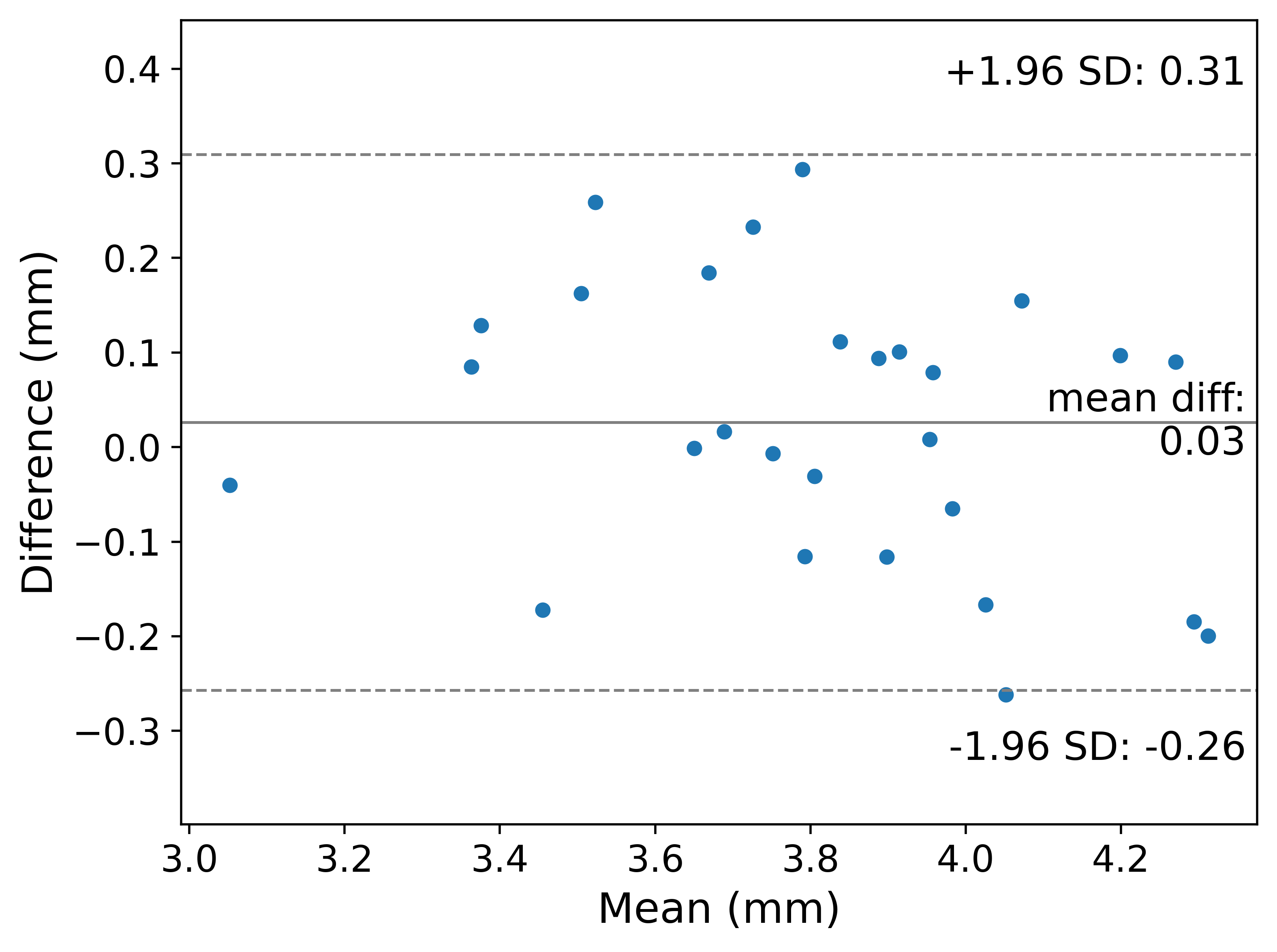}
        \caption{3D U-Net}
        \label{fig:BAunet}
    \end{subfigure}
    \vspace{5pt}
    \caption{Bland-Altman plots showing the difference in transverse thickness between masks generated by SAM 2 (a) and 3D U-Net (b). The difference was calculated by subtracting the ground truth thickness from the generated mask thickness.}
    \vspace{-5pt}
    \label{fig:BlandAltman}
\end{figure}

SAM 2 and U-Net both outperformed SAM 1 in preserving the thickness of the menisci. Bland-Altman plots in Fig.~\ref{fig:BlandAltman} summarise the agreement between the predicted average thickness of the predicted meniscal masks with that of the ground truths. In Bland-Altman plots, the difference between two values is plotted against the mean of the two values \cite{martin_bland_statistical_1986}. Looking at Fig.~\ref{fig:BlandAltman}, it is seen that there is little correlation between the size of the menisci and any under- or over-prediction in thickness. Both SAM 2 and 3D U-Net had a positive average thickness difference, implying that the models overestimated the meniscal thickness, with U-Net overestimating to a lesser degree. These differences were sub-voxel in both cases, so thickness was well-preserved. 

\begin{figure}[t]
\centering
    \begin{subfigure}[h]{0.24\linewidth}
    \centering
        \includegraphics[width = \linewidth]{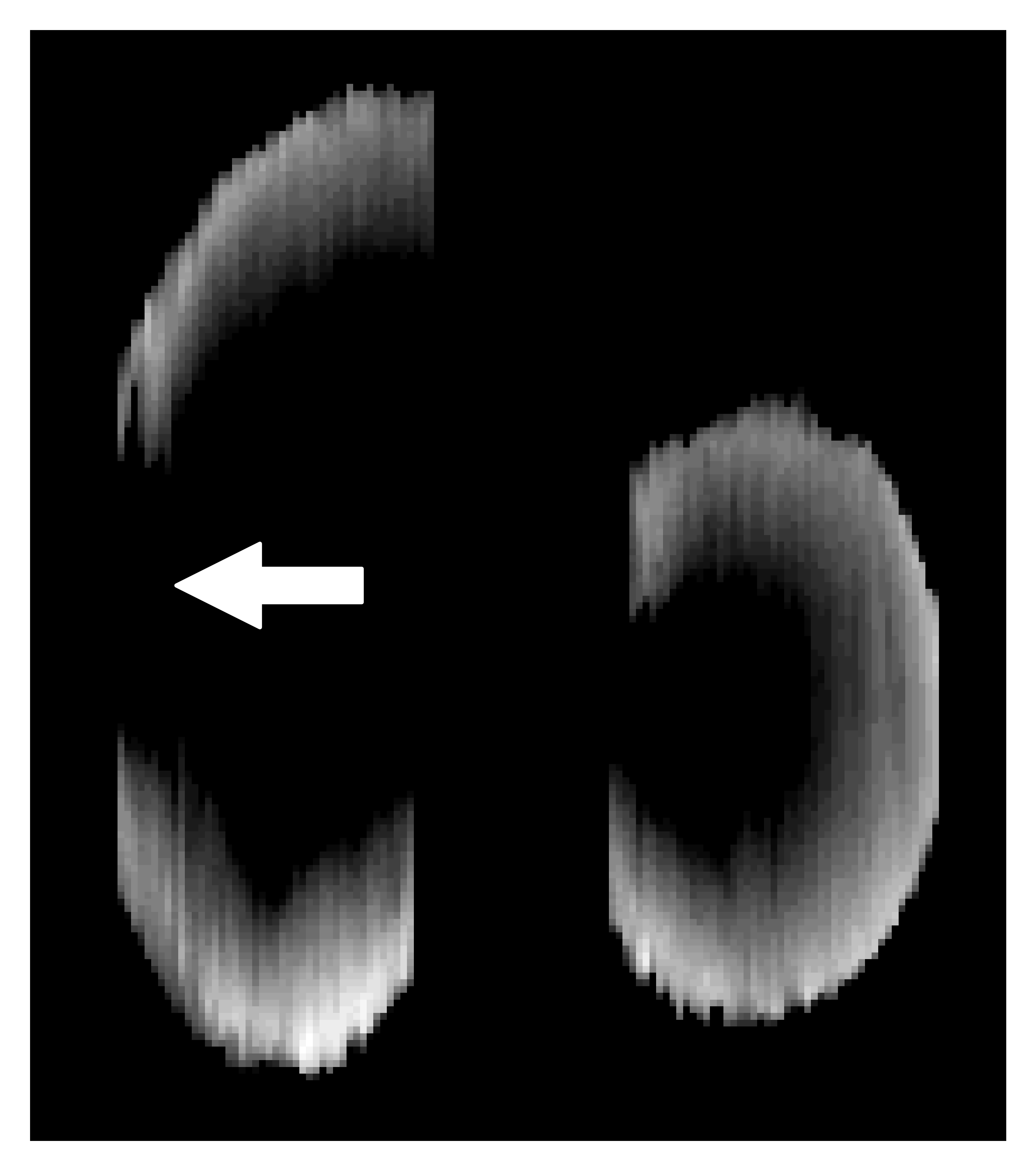}
        \caption{Ground Truth}
        \label{fig:gt_test5}
    \end{subfigure}
    \begin{subfigure}[h]{0.24\linewidth}
    \centering
        \includegraphics[width = \linewidth]{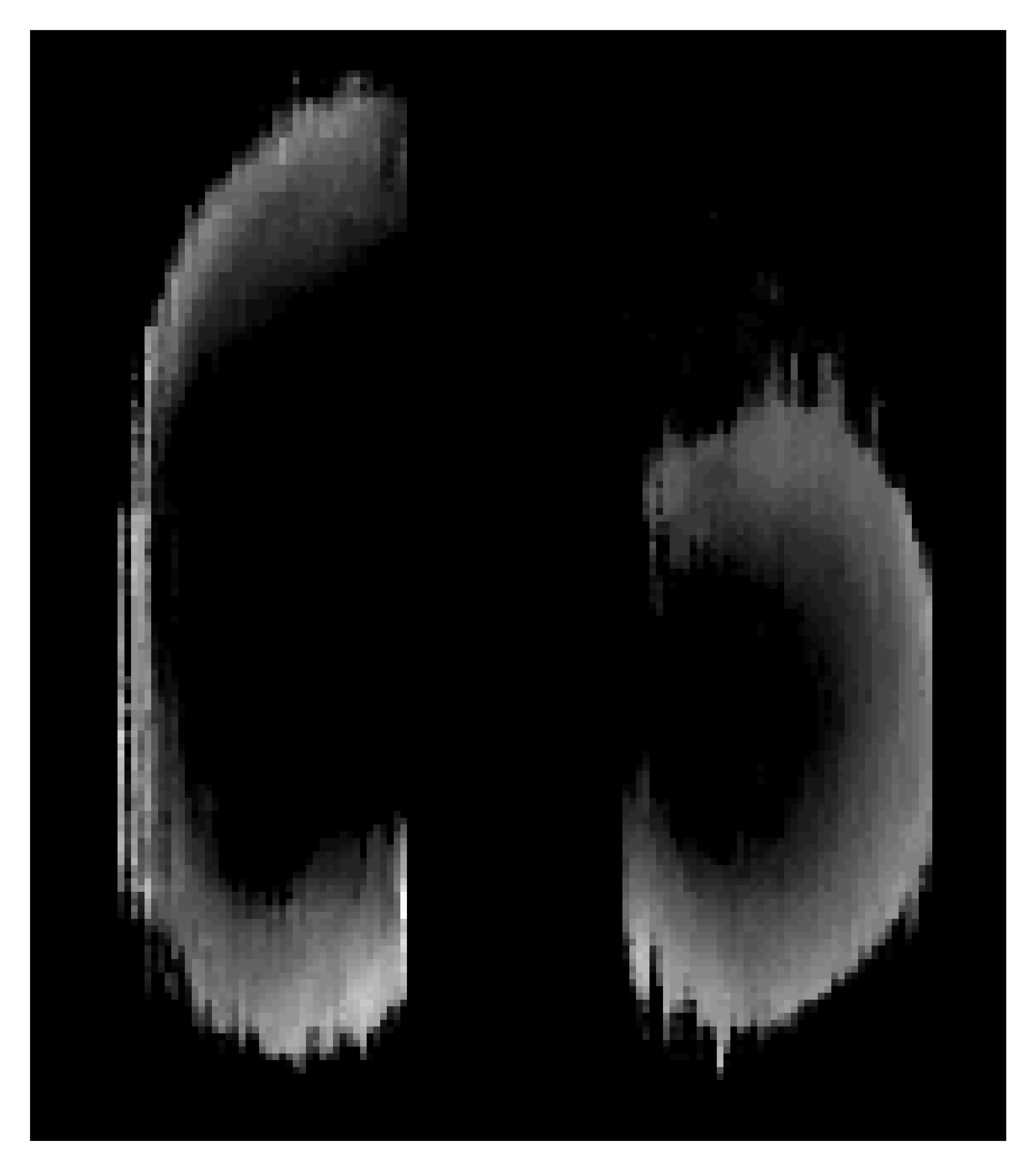}
        \caption{SAM 1}
        \label{fig:sam1_test5}
    \end{subfigure}
    \begin{subfigure}[h]{0.24\linewidth}
    \centering
        \includegraphics[width = \linewidth]{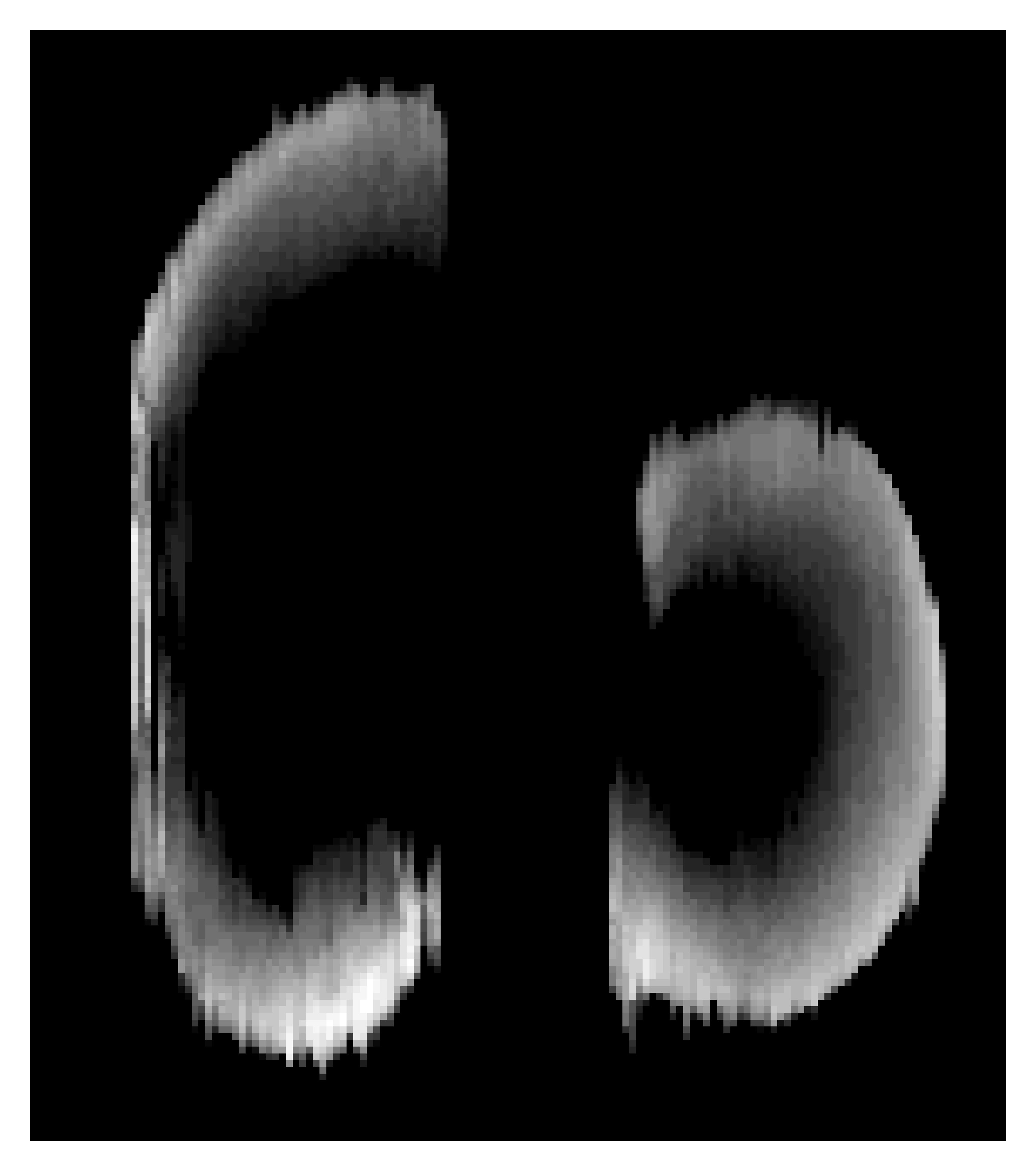}
        \caption{SAM 2}
        \label{fig:sam2_test5}
    \end{subfigure}
    \begin{subfigure}[h]{0.24\linewidth}
    \centering
        \includegraphics[width = \linewidth]{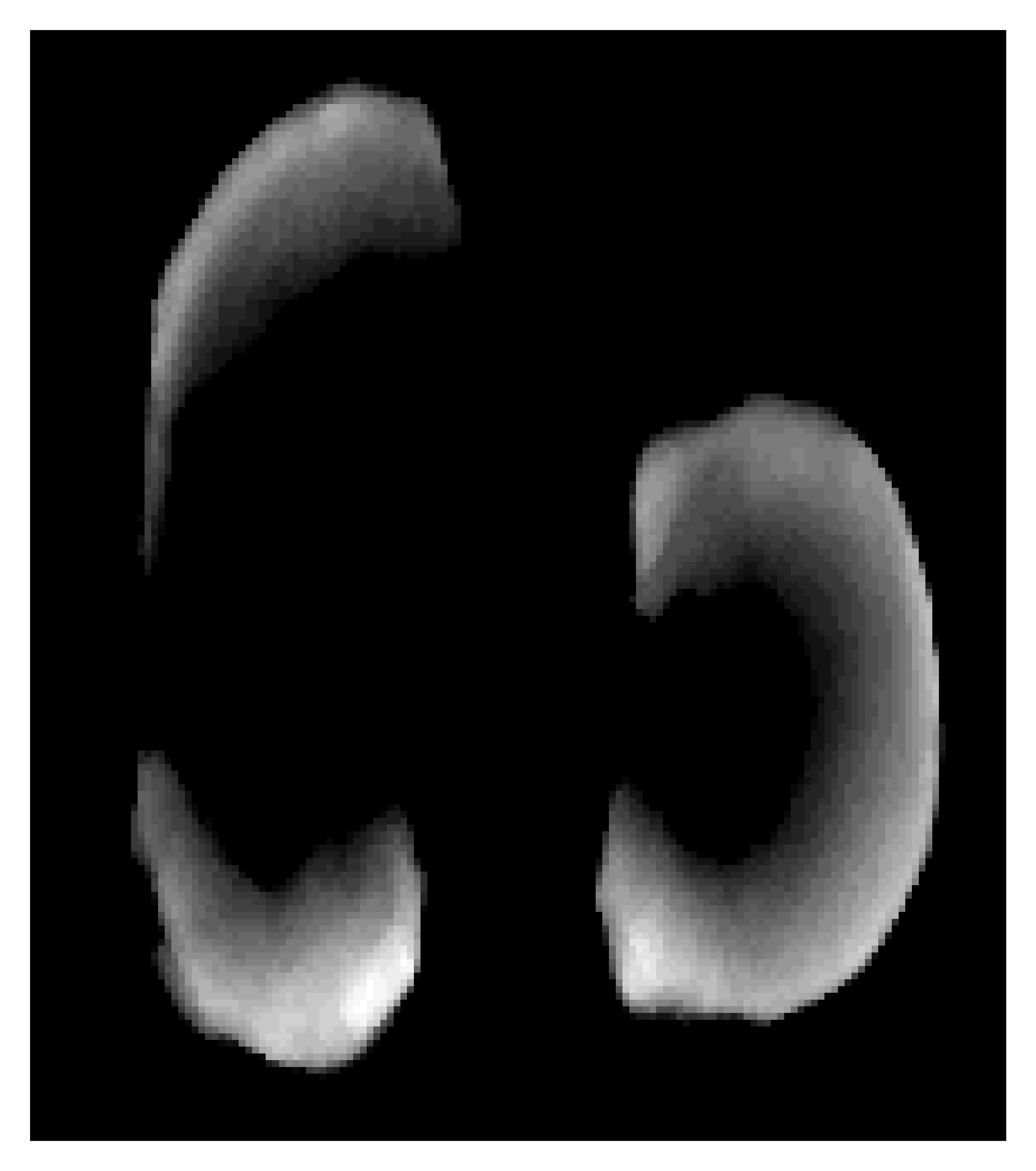}
        \caption{3D U-Net}
        \label{fig:unet_test5}
    \end{subfigure}
    \vfill
    \begin{subfigure}[h]{0.24\linewidth}
    \centering
        \includegraphics[width = \linewidth]{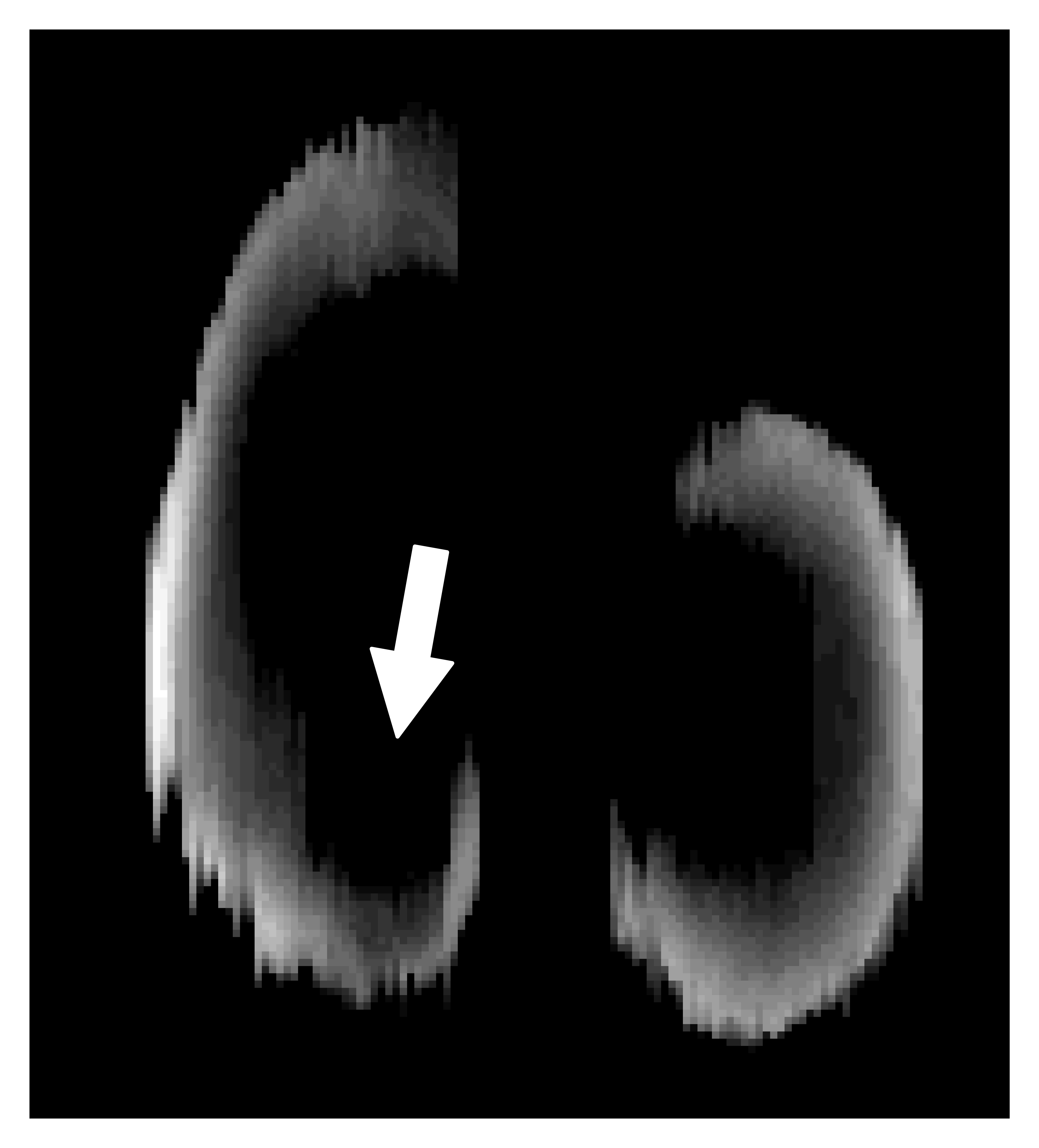}
        \caption{Ground Truth}
        \label{fig:gt_test19}
    \end{subfigure}
    \begin{subfigure}[h]{0.24\linewidth}
    \centering
        \includegraphics[width = \linewidth]{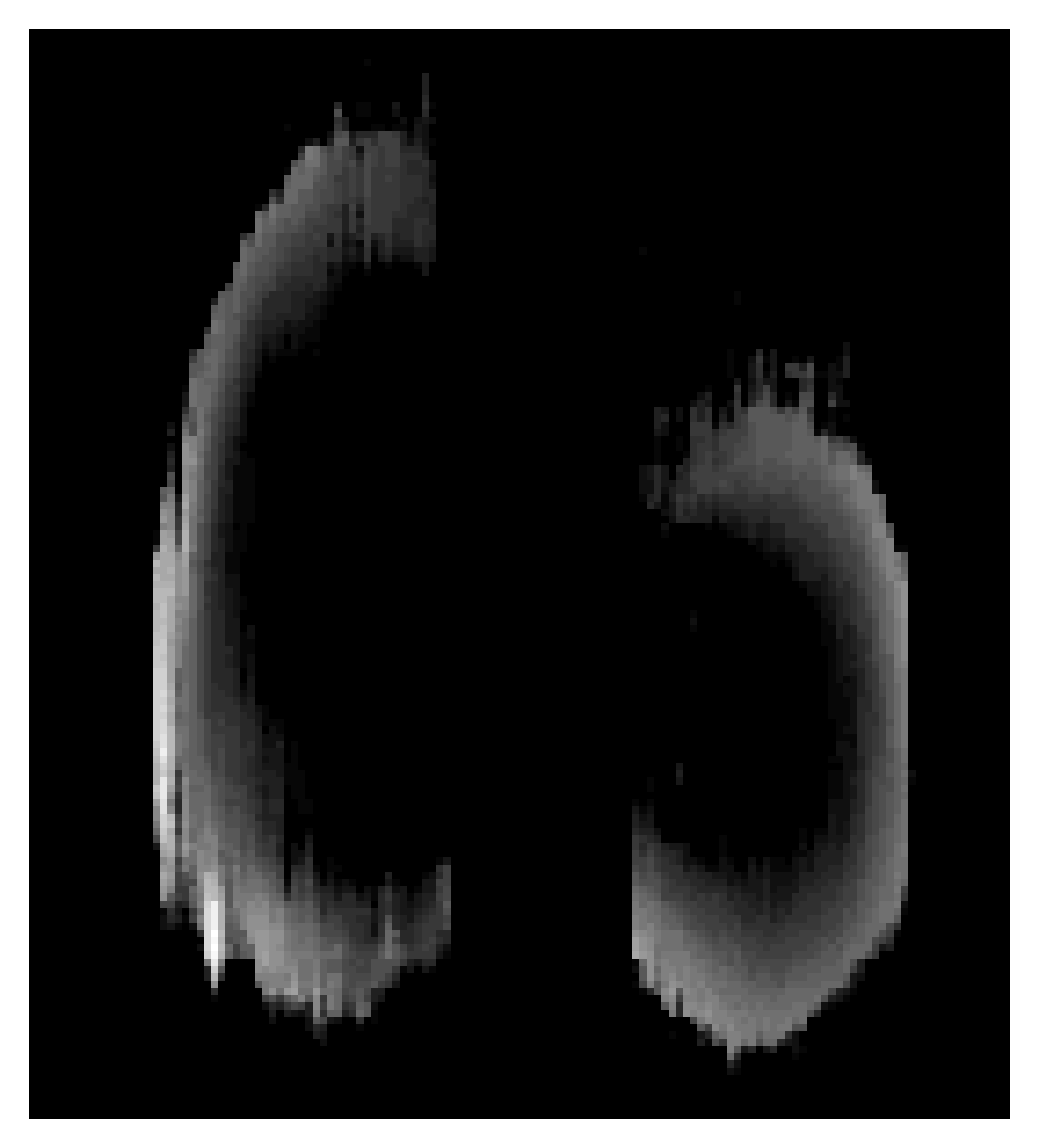}
        \caption{SAM 1}
        \label{fig:sam1_test19}
    \end{subfigure}
    \begin{subfigure}[h]{0.24\linewidth}
    \centering
        \includegraphics[width = \linewidth]{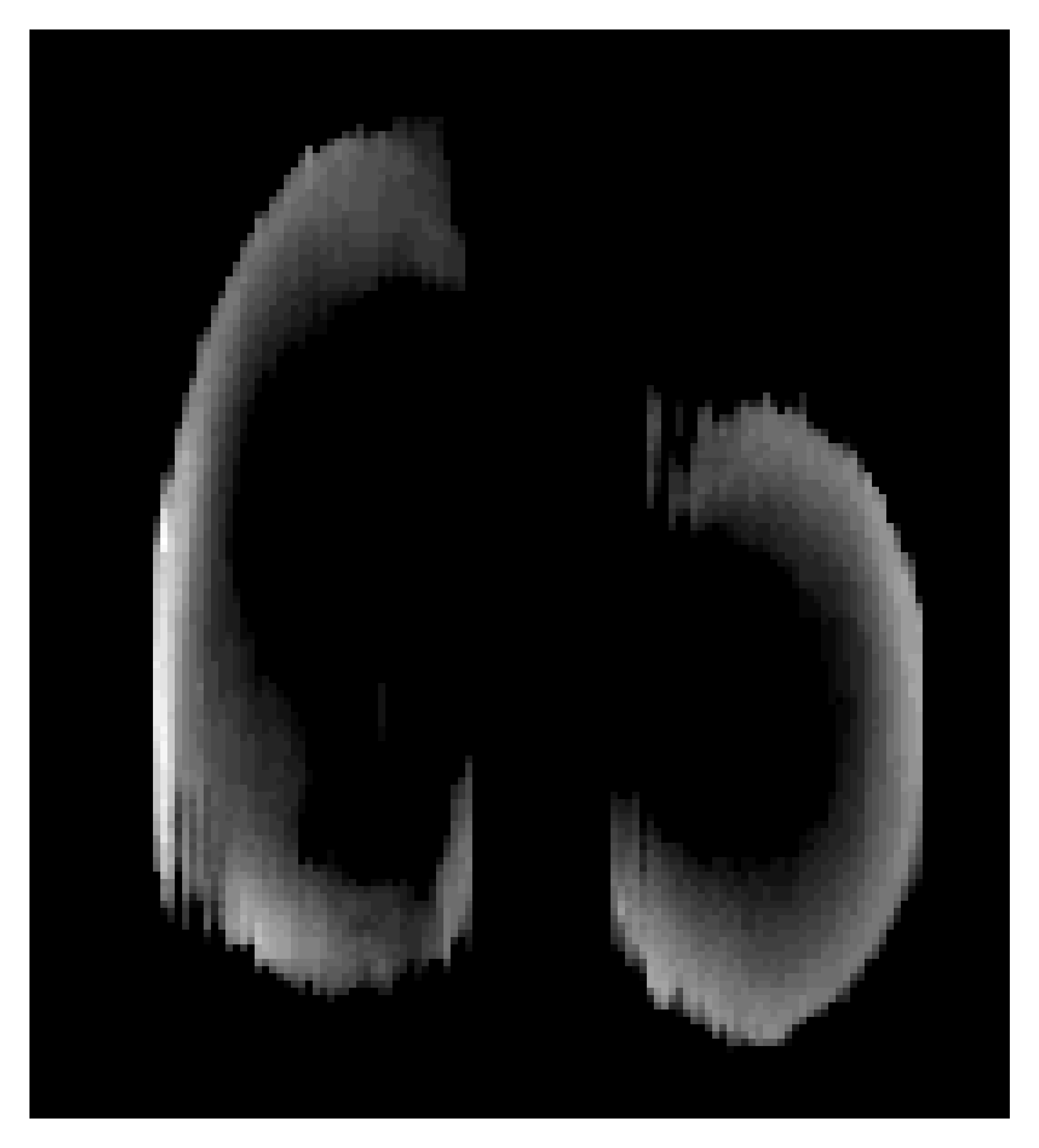}
        \caption{SAM 2}
        \label{fig:sam2_test19}
    \end{subfigure}
    \begin{subfigure}[h]{0.24\linewidth}
    \centering
        \includegraphics[width = \linewidth]{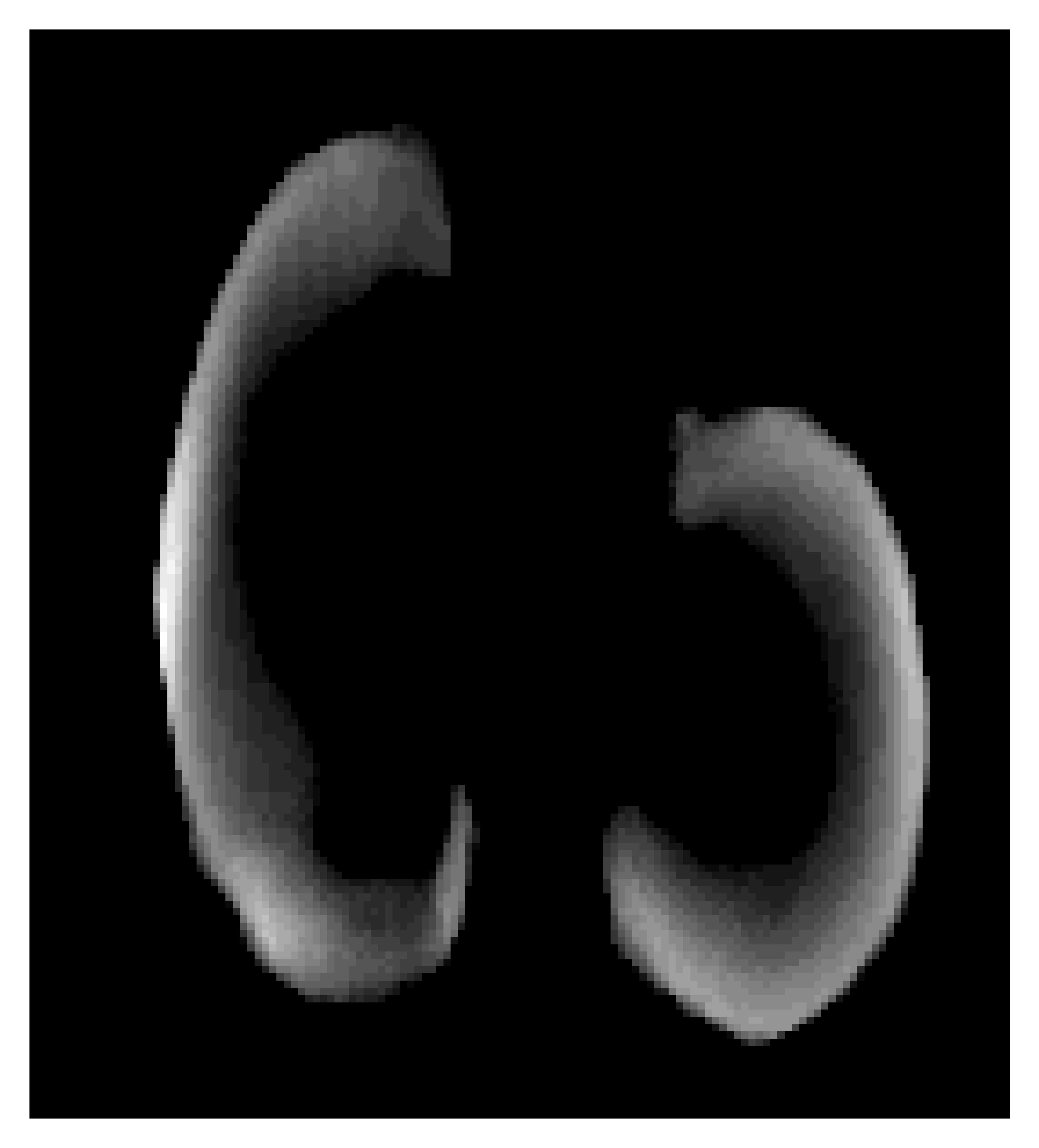}
        \caption{3D U-Net}
        \label{fig:unet_test19}
    \end{subfigure}
    \vspace{5pt}
    \caption{Two atypical examples from the test dataset that visually compare the masks predicted by the segmentation models investigated with the ground truth. In (a), a ground truth mask is shown where the medial meniscus was separated into two (white arrow). (b-d) are the generated masks from SAM 1, SAM 2, and 3D U-Net respectively. (e) shows a test case with what appears to be a partial meniscectomy on the posterior medial meniscal horn (white arrow). (f-h) are again the generated masks from SAM 1, SAM 2, and 3D U-Net respectively. All images in this figure were created by summing the 3D masks through the inferior-superior axis of the body, as if looking down on the mask from above. The brighter a pixel, the thicker the meniscus in this transverse position.}
    \label{fig:examples}
    \vspace{-5pt}
\end{figure}

\subsection{Case Analyses}

Two selected cases from the test set, and the corresponding model predictions, are shown in Fig.~\ref{fig:examples}. One abnormal case in the test set contained a medial meniscus that was fully separated in the middle (a). The models were only exposed to a single case in the train set with a similar morphology. Predicted masks for this test case are shown in (b-d). It can be seen that, despite only being exposed to one similar case in training, 3D U-Net correctly reproduces the two fully-detatched medial meniscus segments. Both SAM configurations struggle to replicate this feature.  Another example was selected due to what appears to be a partial meniscectomy on the posterior horn of the medial meniscus (white arrow in \ref{fig:gt_test19}). The predicted masks for this case indicate that SAM 1 fails to recognise this feature, while SAM 2 and 3D U-Net both do. This demonstrates that training SAM end-to-end improved the ability of the model to extract anomalous morphological features of menisci.

\begin{figure}[t]
\centering
    \begin{subfigure}[h]{0.32\linewidth}
    \centering
        \includegraphics[width = \linewidth]{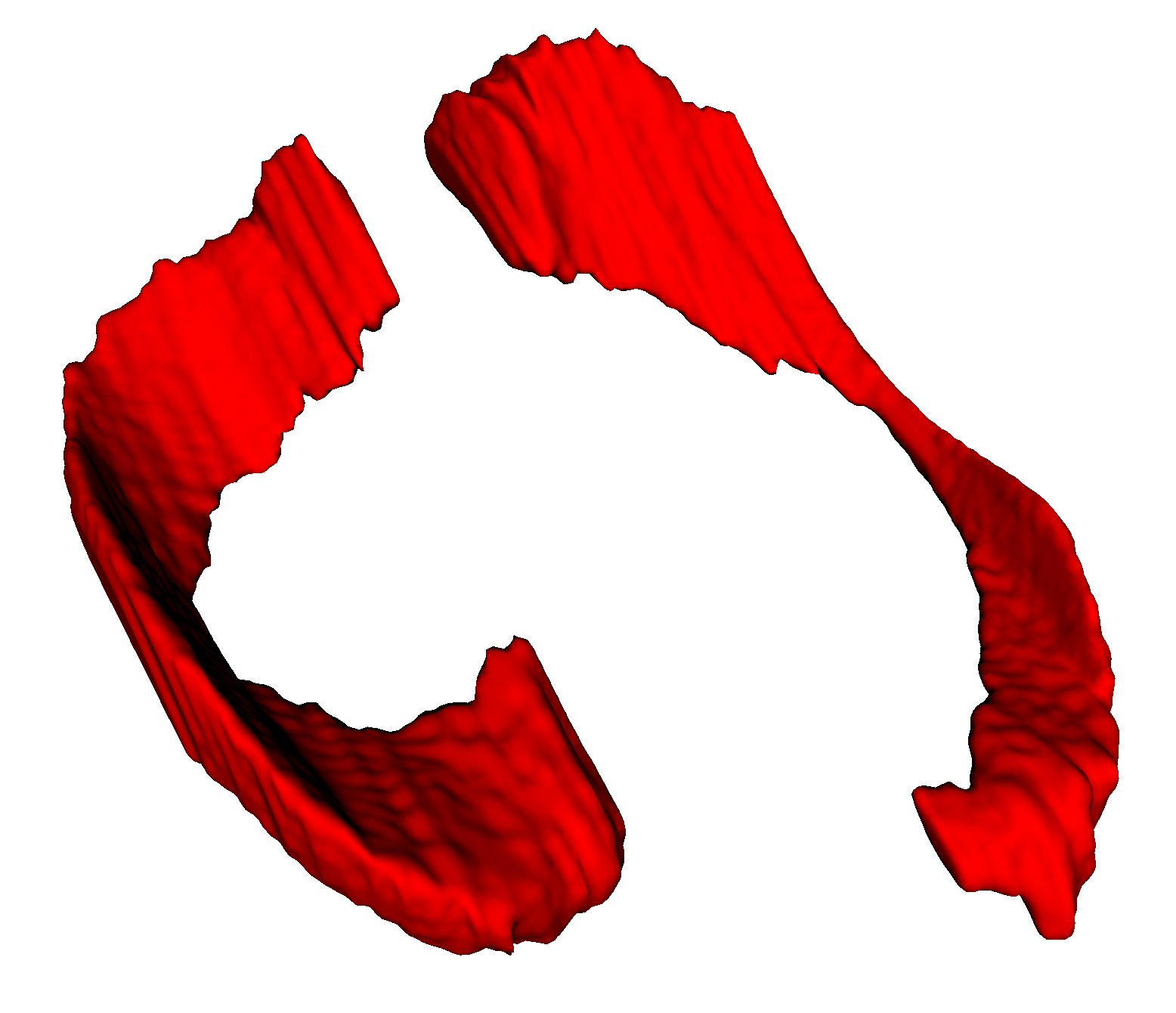}
        \caption{Ground Truth}
        \label{fig:gt_mask}
    \end{subfigure}
    \begin{subfigure}[h]{0.32\linewidth}
    \centering
        \includegraphics[width = \linewidth]{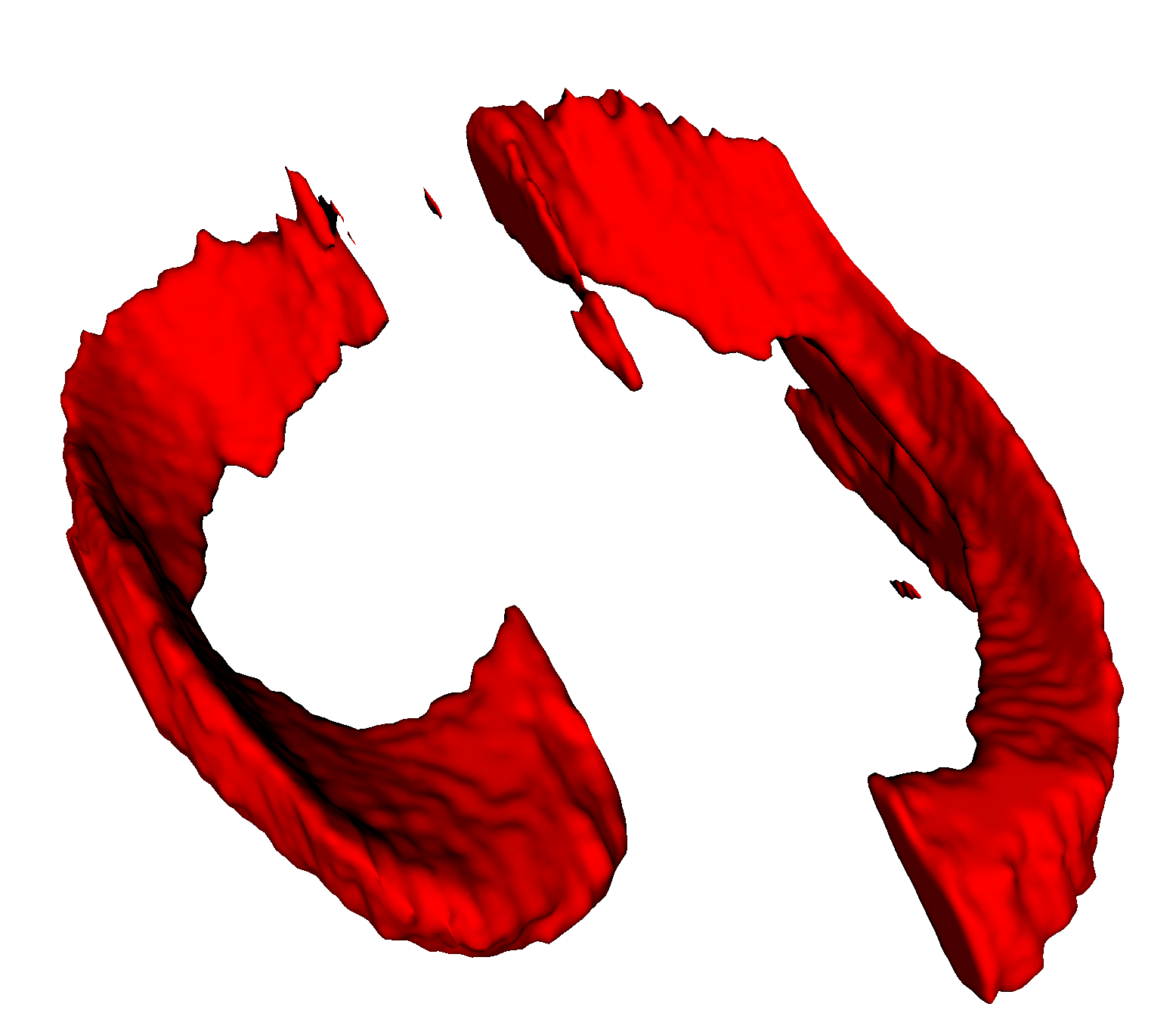}
        \caption{SAM 2}
        \label{fig:sam2_maskpred}
    \end{subfigure}
    \begin{subfigure}[h]{0.32\linewidth}
    \centering
        \includegraphics[width = \linewidth]{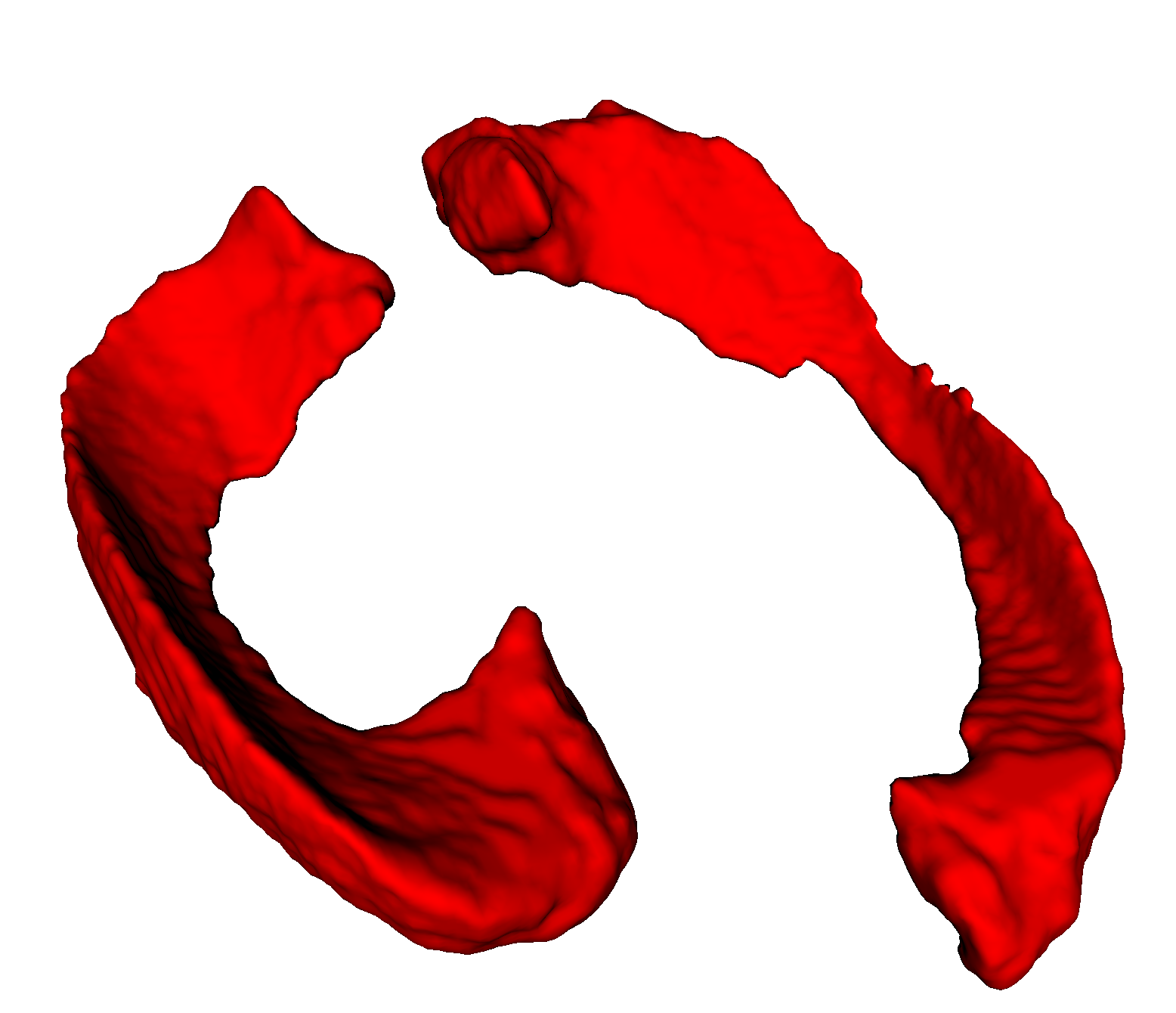}
        \caption{3D U-Net}
        \label{fig:unet_maskpred}
    \end{subfigure}
    \vspace{5pt}
    \caption{Surface mesh representations of the worst-performing (Dice score) predicted masks of 3D U-Net and end-to-end trained SAM (SAM 2) from the test knee MRI set, shown alongside the corresponding ground truth. The meshes contain both the lateral and medial menisci, which are seen on the left and right of each subfigure.}
    \vspace{-10pt}
    \label{fig:masks}
\end{figure}

Fig.~\ref{fig:masks} displays surface mesh representations of masks generated from a test set image by SAM 2 and U-Net, along with the corresponding ground truth. The 3D meshes were produced using ITK-SNAP. Both SAM 2 and U-Net scored the lowest Dice coefficient on this image, so the meshes are a good example of where SAM 2 and U-Net fail to accurately perform meniscus segmentation. The first feature to highlight is that SAM 2 fails to replicate the narrowing near the middle of the medial meniscus. U-Net does much better at reproducing the general geometry of the ground truth. SAM 2 also struggles to keep the menisci as a contained volume, with flecks of unattached positive predictions seen in Fig.~\ref{fig:sam2_maskpred}. U-Net did not suffer from this problem, which likely stemmed from SAM 2 lacking 3D contextual information unlike the former.

SAM configurations often outputted masks containing small isolated islands of positive prediction (e.g. Fig.~\ref{fig:sam2_maskpred}), struggling to output consistent intact volumes. Through performing connected component analysis on the ground truths and generated masks, it was seen that predictions from SAM 1 and 2 contained an average of $46.9$ and $10.2$ components respectively, compared to the ground truths which contained an average of $2.1$. In contrast, masks generated by 3D U-Net contained very few regions disconnected from the main segmentation bodies (average number of connected components of $2.3$). This is desirable, because less post-processing would be required if the generated masks were to be analysed geometrically.

Figures~\ref{fig:examples}~and~\ref{fig:masks} show that the masks generated by 3D U-Net are smoother, whereas the texture of the SAM masks more closely matches the ground truth. This could be due to the ground truths being annotated slice-wise leading to staircase artifacts that are not faithful 3D representations of the menisci. In this case, the Dice score may be punishing the 3D U-Net model for generating smoother masks than the ground truths, which may actually more closely resemble the true 3D structure of the menisci. There was concern that this smoothing effect might result in 3D U-Net smoothing out smaller features, but the Hausdorff distance results show that the smoothing is not compromising the model's ability to match the meniscus geometry.

One problem with the data set used in this study was that, apart from two cases in the training data and one in the test data, all menisci were fully intact. When menisci are highly degenerate, they may no longer be two semi-lunar structures. Segmenting menisci once they have broken down into multiple pieces would be a far more challenging task, but there was little scope here to monitor performance on such cases. Another issue was that the ground truths were annotated by a single expert, so prone to more bias and error than if multiple segmenters had resolved conflicts between themselves.

\section{Conclusion}

Despite its generalisability, SAM, when only fine-tuning the decoder, performed significantly worse than a 3D U-Net when segmenting the menisci from MR images (Dice score $0.81\pm0.03$ compared to $0.87\pm0.03$). When training SAM end-to-end, comparable results were obtained (Dice score $0.87\pm0.03$). This demonstrates that fine-tuning the SAM encoder, to better extract task-relevant features, can compete with state-of-the-art performance ($0.88\pm0.03$). However, the Hausdorff Distance showed that SAM was inferior to 3D U-Net in preserving the spatial features of the menisci, showing that it may not be suitable for deriving meaningful biomarkers from the menisci for monitoring the progression of OA. These weaknesses of SAM may also be relevant for other medical image segmentation tasks involving fine anatomical structures with low contrast and unclear boundaries.

\section{Future Work}

In the data set used, each patient had two MR scans at different points in time. For patients with noticeable changes between the two scans, it would be beneficial to analyse the ability of models to identify these changes. For example, if the thickness has changed over time, this should be mirrored in the generated masks.

Once a suitable model has been selected, automated segmentation of menisci could be performed for the entire OAI cohort. Geometric analysis of the generated masks, in combination with patients' clinical data, has the potential to reveal important OA biomarkers, both for onset and progression.


\section*{Acknowledgments}

We would like to thank the OAI and its participants for creating this publicly available data set, and Dr Akshay Chaudhari (Stanford University) for providing the subset of the OAI used for the IWOAI 2019 challenge. This work made use of time on both ARC4, part of the High Performance Computing facilities at the University of Leeds, and Tier 2 HPC facility JADE2 (funded by EPSRC, EP/T022205/1). This work was funded by EPSRC (EP/S024336/1). Code is available at: \url{https://github.com/oliverjm1/BMVC_menisc_seg}.

\bibliography{references.bib, extrarefs.bib}
\end{document}